\documentclass[aps,prb,twocolumn,groupedaddress,showkeys,showpacs]{revtex4}

\usepackage{multirow,amssymb,amsbsy,amsmath}
\usepackage{epsfig}

\newcommand{\op}[1]{%
    \fontdimen12\textfont3=2pt\fontdimen12\scriptfont3=1.4pt%
    \!\null\mathop{\vphantom{#1}\smash{#1}}\limits_{\sim}\null\!}
\newcommand{\xref}[1]{\protect\ref{#1}}
\newcommand{\figref}[1]{Fig.~\protect\ref{#1}}

\newcommand {\mofe} {\{$\textrm{Mo}_{72}\textrm{Fe}_{30}$\}}
\newcommand {\mov} {\{$\textrm{Mo}_{72}\textrm{V}_{30}$\}}
\newcommand {\cula} {\{$\textrm{Cu}_{12}\textrm{La}_{8}$\}}
\newcommand{\pp}[2]{\frac{\partial \, {#1}}{\partial \, {#2}}}

\newcommand{\Bsat}{B_{\text{sat}}}

\begin{document}

\title{Enhanced magnetocaloric effect in frustrated magnetic
  molecules with icosahedral symmetry}

\author{J\"urgen Schnack}
\email{jschnack@uni-bielefeld.de}
\affiliation{Universit\"at Bielefeld, Fakult\"at f\"ur Physik,
  Postfach 100131, D-33501 Bielefeld, Germany}
\author{Reimar Schmidt}
\email{reimar.schmidt@phyisk.uni.magdeburg.de}
\author{Johannes Richter}
\email{Johannes.Richter@physik.uni-magdeburg.de}
\affiliation{Institut f\"ur Theoretische Physik, Universit\"at Magdeburg,
P.O. Box 4120, D-39016 Magdeburg, Germany}

\date{\today}

\begin{abstract}

We investigate the magnetocaloric properties of certain
antiferromagnetic spin systems that have already been or very
likely can be synthesized as magnetic molecules.  It turns out
that the special geometric frustration which is present in
antiferromagnets that consist of corner-sharing triangles leads
to an enhanced magnetocaloric effect with high cooling rates in
the vicinity of the saturation field. These findings are
compared with the behavior of a simple unfrustrated spin ring as
well as with the properties of the icosahedron. To our surprise,
also for the icosahedron large cooling rates can be achieved but
due to a different kind of geometric frustration.

\end{abstract}

\maketitle

\section{Introduction}
\label{sec-1}

Antiferromagnetic finite-size spin systems with icosahedral
symmetry constitute very interesting frustrated materials with
rather unusual magnetic properties. Among such properties are
jumps to the saturation magnetization in the cuboctahedron and
the icosidodecahedron\cite{SSR:EPJB01,SSR:JMMM05} as well as
metamagnetic phase transitions at zero temperature for instance
in the icosahedron and
dodecahedron.\cite{CoT:PRL92,SSS:PRL05,Kon:PRB05,Kon:PRB06} Some
of these properties, for instance the large magnetization jump
to saturation, are as well present in the Kagome or other
lattice antiferromagnets.\cite{SHS:PRL02} The finite-size systems, nevertheless,
have the advantage that due to their smallness many properties
can be investigated (numerically) exactly with possible benefits
for our knowledge about infinitely extended frustrated spin
systems.  In this article we investigate the magnetocaloric
properties of certain spin clusters with icosahedral symmetry
that turn out to be interesting as well.

Magnetocalorics has a long tradition especially in connection
with cooling by adiabatic demagnetization. The first successful
attempts to reach the sub-Kelvin region date back more than 70
years.\cite{GiM:PR33} It is equally well possible to extend such
an adiabatic process to a full Carnot-cycle and thus use the
magnetization work for magnetic refrigeration.\cite{PeG:JMMM99}
In the past paramagnetic salts have been the working medium in
both cases which limits the cooling rate to be not more than
2~K/T. Nowadays gadolinium compounds such as
Gd$_5$Ga$_5$O$_{12}$ or Gd$_5$Si$_2$Ge$_2$ are known to be very
efficient refrigerant materials.\cite{PeG:PRL97,Zhi:PRB03}

A unified explanation of a magnetocaloric effect that is
enhanced compared to a paramagnet is provided by the observation
that the cooling rate assumes extreme values close to
configurations with a large excess entropy. This can happen at
certain phase transitions such as the first order
[ferromagnetic(I) $\leftrightarrow$ ferromagnetic(II)] phase
transition observed in Gd$_5$Si$_2$Ge$_2$
compounds,\cite{PeG:PRL97} at magnetic field driven transitions
across a quantum critical point,\cite{ZGR:PRL03,ZhH:JSM04} or at
special values of the magnetic field where many ground state
Zeeman levels are
degenerate.\cite{ZhH:JSM04,DeR:PRB04,ZhT:IKYS05}

In this article we demonstrate that an enhanced magnetocaloric
effect should be observable in certain highly frustrated
magnetic molecules of icosahedral symmetry. We discuss the
cuboctahedron\cite{BGG:JCSDT97} and the
icosidodecahedron\cite{MSS:ACIE99} which are already synthesized
Archimedian solids as well as the icosahedron which in its full
symmetry could not yet be achieved
chemically. \cite{BGG:JCSDT97,BGH:JCSDT97,TEM:CEJ06} Some of the
aspects of our investigation have been previously discussed in
connection with the classical\cite{Zhi:PRB03} as well as in the
quantum version\cite{ZhT:IKYS05} of the Kagome lattice
antiferromagnet and some one-dimensional
antiferromagnets.\cite{ZhH:JSM04,DeR:EPJB06} We also like to
mention that magnetocaloric studies have been carried out in the
field of molecular magnetism recently, but were mainly focused
on low-field
behavior.\cite{MSS:JMMM92,BMT:JAP94,THB:APL00,ZWZ:PRL01,WKS:PRL02,TBH:JPCM03,AGC:APL04,ECG:APL05}

The article is organized as follows. In Sec.~\xref{sec-2} we
discuss the basics of magnetocalometry, whereas in
Sec.~\xref{sec-3} we present the magnetocaloric properties of
icosahedral bodies and compare them with those of a
non-frustrated spin ring of the same size. The paper closes with
a summary.

\section{Basic thermodynamics}
\label{sec-2}

\subsection{Heisenberg Model}
\label{sec-2-1}

The spin systems discussed in this article are modeled by an
isotropic Heisenberg Hamiltonian augmented with a Zeeman term,
i.e., 
\begin{eqnarray}
\label{E-2-1}
\op{H}
&=&
-
\sum_{u, v}\;
J_{uv}\,
\op{\vec{s}}(u) \cdot \op{\vec{s}}(v)
+
g \mu_B B \op{S}_z
\ .
\end{eqnarray}
$\op{\vec{s}}(u)$ are the individual spin operators at sites
$u$, $\op{\vec{S}}=\sum_u\op{\vec{s}}(u)$ is the total spin
operator, and $\op{S}_z$ its $z$-component along the homogeneous
magnetic field. $J_{uv}$ are the matrix elements of the
symmetric coupling matrix. A negative value of $J_{uv}$
corresponds to antiferromagnetic coupling. For the symmetric
polytopes discussed in the following an antiferromagnetic
nearest-neighbor exchange of constant size $J$ is assumed.

\subsection{Magnetocaloric effect}
\label{sec-2-2}

\begin{figure}[ht!]
\centering
\includegraphics[clip,width=55mm]{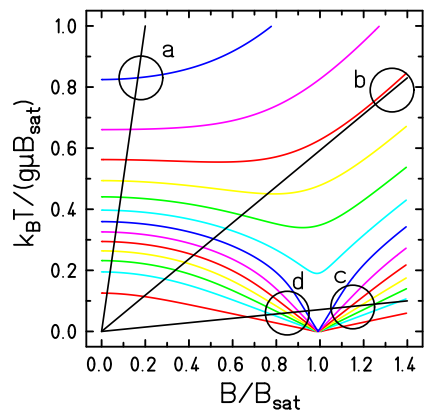}
\caption{(Color online) The curves show isentropes of the
  antiferromagnetically coupled dimer of two spins with
  $s=1/2$. The straight lines represent isentropes of a
  paramagnet. Compared to a paramagnet the cooling rate of the
  antiferromagnetic dimer can be smaller (a), about the same
  (b), much bigger (c), or even negative (d).}
\label{F-A}
\end{figure}

The magnetocaloric effect consists in cooling or heating of a
magnetic system in a varying magnetic field. Some basic
thermodynamics yields the adiabatic (i.e. isentropic,
$S=const.$) temperature change as function of temperature and
applied magnetic field,
\begin{eqnarray}
\label{mce-2-1}
\left(
\pp{T}{B}
\right)_S
&=&
-
\frac{T}{C(T,B)}
\left(\pp{S}{B}\right)_T
\ .
\end{eqnarray}
This rate is also called cooling rate. $C(T,B)$ is the
temperature- and field-dependent heat capacity of the system.
For a paramagnet this rate is simply\cite{Zhi:PRB03}
\begin{eqnarray}
  \label{mce-2-2}
\left(
\pp{T}{B}
\right)_S^{\text{para}}
&=&
\frac{T}{B}
\ .
\end{eqnarray}
This situation changes completely for an interacting spin
system.  Depending on the interactions the adiabatic cooling
rate $\pp{T}{B}$ can be smaller or bigger than the paramagnetic
one and even change sign, i.e. one would observe heating during
demagnetization and cooling during
magnetization.\cite{WKS:PRL02,ZhH:JSM04} For the purpose of
clarity this will be shortly illuminated with the help of an
antiferromagnetically coupled dimer of two spins with $s=1/2$,
where a singlet constitutes the ground state and a triplet the
excited state. Following \eqref{mce-2-1} one notices that in the
vicinity of the magnetic field $B_c$, where the lowest triplet
level crosses the singlet, the entropy changes drastically at
low temperatures due to the fact that at the crossing field the
ground state is degenerate whereas elsewhere it is not. This
behavior is displayed in \figref{F-A}, where below and above the
crossing field the cooling rate, i.e. the slope of the
isentropes, assumes large values.

\section{Magnetocaloric effect in magnetic molecules}
\label{sec-3}

Regarding the use as a magnetic refrigerant material magnetic
molecules possess several advantageous properties. They can be
synthesized in a great variety of structures and they can host
various paramagnetic ions. Very often they also do not interact
magnetically with each other in a bulk sample due to large
distances between the magnetic centers of different molecules
that are provided by extended ligands, which means that the
magnetic properties of a single molecule can be assigned to the
macroscopic sample. If it would be possible to obtain structures
with exhibit extraordinarily large ground-state degeneracies at
certain magnetic fields one could exploit these materials for
very efficient magnetization cooling.

Every antiferromagnetically coupled spin system exhibits ground
state level crossings such as the aforementioned spin dimer. The
cooling rate at such a crossing can assume large values, but it
turns out that it is possible to even increase the rate in a
special class of highly frustrated magnetic molecules of
icosahedral symmetry. This will be discussed for the
cuboctahedron, which is chemically realized as a \cula\
molecule,\cite{BGG:JCSDT97} and for the icosidodecahedron, which
is chemically realized as a \mofe\ molecule\cite{MSS:ACIE99} and
a \mov\ molecule\cite{MTS:AC05} as well as for the not yet
synthesized icosahedron. The behavior of these frustrated spin
systems is compared with the behavior of an antiferromagnetic,
but not frustrated spin ring with $N=12$ sites. All results are
obtained by means of numerical diagonalization.

The geometric structures of the discussed bodies can be
visualized for instance at the following
Refs.~\onlinecite{SSR:EPJB01,SSR:JMMM05,mathworld}.

The cuboctahedron is one of the smallest antiferromagnetic spin
systems that can host independent localized
magnons.\cite{SSR:EPJB01,SHS:PRL02,SSR:JMMM05} These localized
states are intimately connected with an enhanced degeneracy of
energy levels and -- in extended spin systems such as the Kagome
lattice antiferromagnet -- with the appearance of flat bands. In
addition, the possibility to arrange several independent magnons
on the (finite) spin lattice results in a linear dependence of
the minimal energy $E_{\text{min}}(M)$ on the total magnetic
quantum number $M$. Therefore, at the saturation field $\Bsat$ a
massive degeneracy of ground state levels can be achieved.

\begin{figure}[ht!]
\centering
\includegraphics[clip,width=55mm]{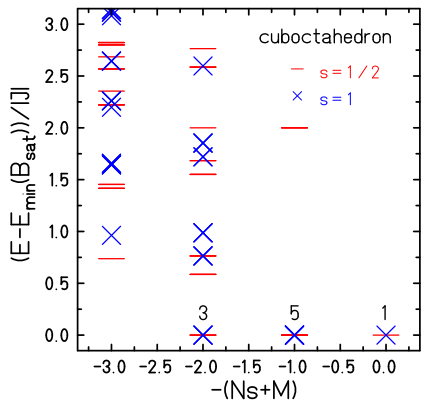}
\caption{(Color online) Low-lying energy levels of the
  cuboctahedron with $s=1/2$ (dashes) and $s=1$ (x-symbols) at
  the saturation field $\Bsat$. The attached numbers give the
  multiplicity $d_M$ of each $M$-level.}
\label{F-B}
\end{figure}

Figure~\xref{F-B} shows that even in a system as small as the
cuboctahedron the ground state multiplicity at the saturation
field can reach a rather large value, in this case of nine
independent of spin quantum number.\cite{SSR:JMMM05,SSR:EPJB01} Thus, the
entropy assumes a non-zero value at zero temperature of
$S_0=S(T=0,B=\Bsat)=k_B \ln(9)$. All isentropes with $S\le S_0$
therefore arrive at the phase space point ($T=0, B=\Bsat$).

\begin{figure}[ht!]
\centering
\includegraphics[clip,width=85mm]{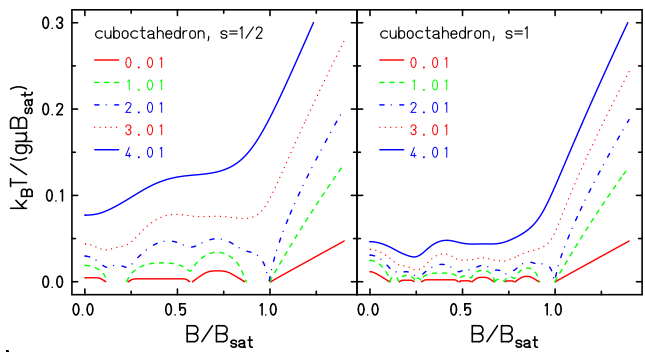}
\caption{(Color online) Isentropes of the cuboctahedron with $s=1/2$ and
  $s=1$. The entropies here and in the following are given in
  units of Boltzmann's constant $k_B$; from the upper left to
  the lower right the values of $S/k_B$ are 0.01, 1.01, 2.01,
  3.01, and 4.01, respectively. $\Bsat=(12 s |J|)/(g\mu_B)$.}
\label{F-C}
\end{figure}

Figure~\xref{F-C} displays various isentropes of the
cuboctahedron both for $s=1/2$ (l.h.s.) as well as for $s=1$
(r.h.s.). Both temperature and magnetic field are normalized to
the saturation field here and in the following. One clearly sees
that the low-entropy curves, $S\le S_0$, approach the $B$-axes
in a rather universal way independent of the magnitude of the
spin $s$. Close to the saturation field the slope of the
isentropes, i.e. the cooling rate, can assume large values.
Below the saturation field the isentropes remain rather flat,
i.e. the temperature does not increase again when going to
$B=0$. This is of course an important property because otherwise
the system would heat up again when switching off the field. 

\begin{figure}[ht!]
\centering
\includegraphics[clip,width=85mm]{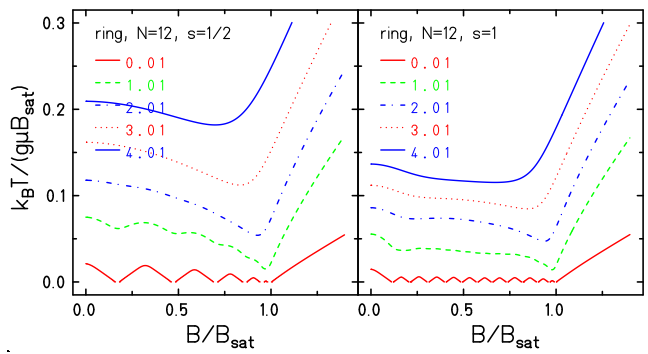}
\caption{(Color online) Isentropes of the a bipartite,
  i.e. non-frustrated spin ring with $N=12$ spins $s=1/2$ and
  $s=1$. $\Bsat=(8 s |J|)/(g\mu_B)$.} 
\label{F-D}
\end{figure}

The non-frustrated ring system, \figref{F-D}, does not possess
the later property nor does it exhibit large cooling rates at
low temperatures close to the saturation field. The second
property is easily understood because the massive degeneracy at
the saturation field does not occur in an antiferromagnetic
ring. Therefore, only isentropes with $S<k_B \ln(2)$ approach the
field axis. The first property, however, is related to the
overall structure of the low-lying levels. In finite bipartite
systems the levels are arranged in rotational
bands\cite{ScL:PRB01,Wal:PRB01} which means that in each sector
of total magnetic quantum number $M$ the excited states are
separated from the ground state in the respective sector by a
non-vanishing gap. Thus at lower magnetic fields a certain
entropy can only be realized by populating higher-lying levels,
i.e. by acquiring a higher temperature.

\begin{figure}[ht!]
\centering
\includegraphics[clip,width=85mm]{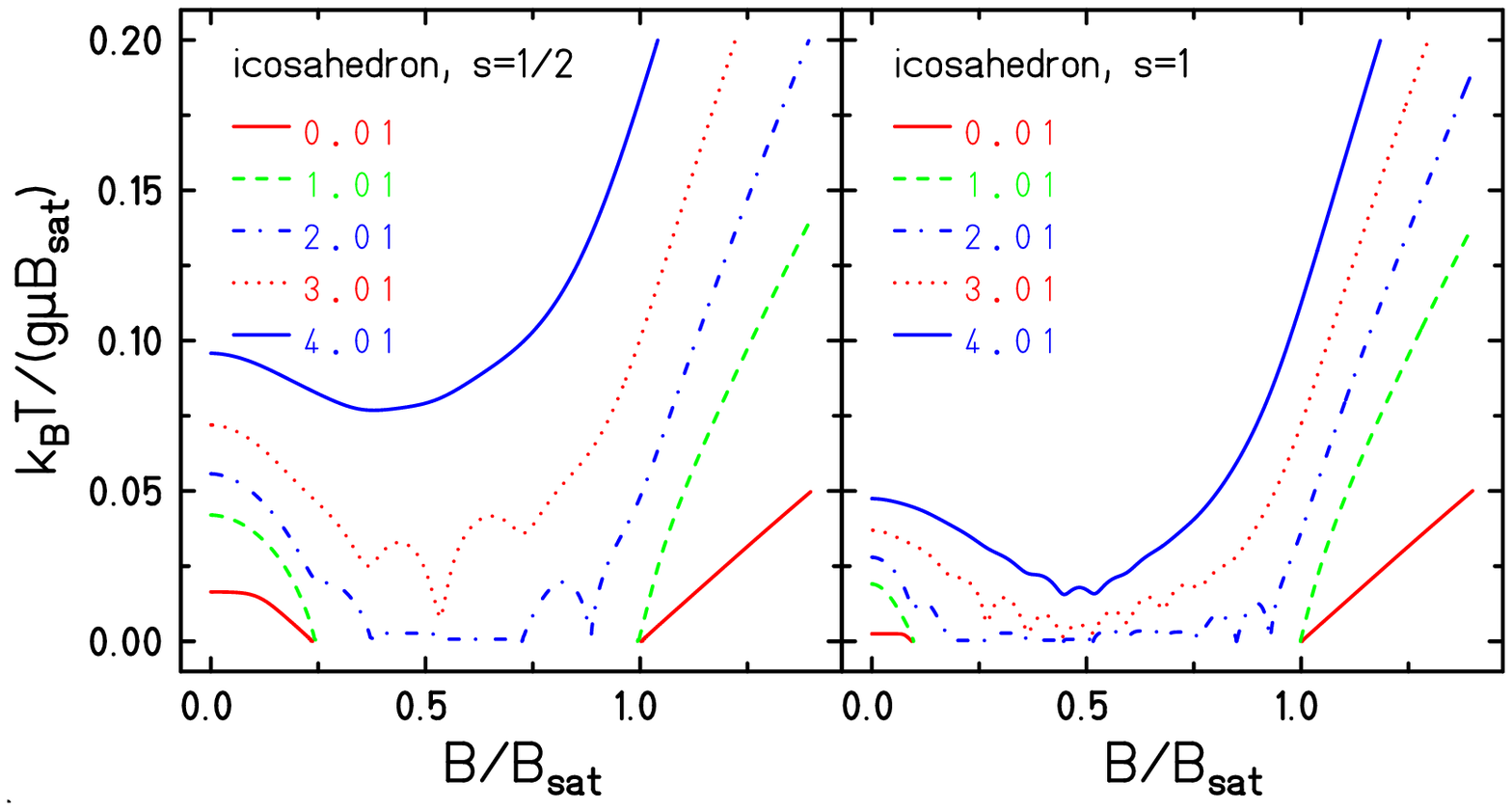}
\caption{(Color online) Isentropes of the icosahedron with
  $s=1/2$ and $s=1$. $\Bsat=(14.472 s |J|)/(g\mu_B)$.}
\label{F-E}
\end{figure}

The behavior of the icosahedron is intermediate. The
antiferromagnetic icosahedron is also a geometrically frustrated
spin system but differs from the cuboctahedron in some
aspects. Each spin has five nearest neighbors (cuboctahedron --
four), and the structure consists of edge-sharing triangles
instead of corner-sharing triangles as for the
cuboctahedron. Therefore, the icosahedron does not possess
independent one-magnon states and consequently a similar
degeneracy at the saturation field cannot be expected, compare
\figref{F-E}. 

\begin{figure}[ht!]
\centering
\includegraphics[clip,width=55mm]{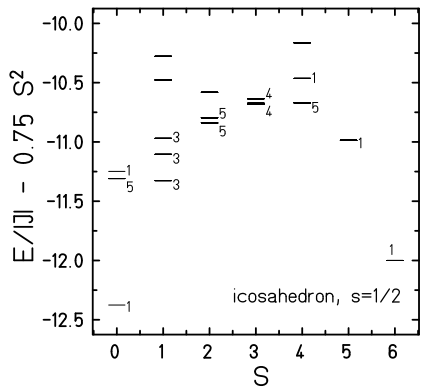}
\caption{Low-lying energy levels of the icosahedron with
  $s=1/2$. The attached numbers give the multiplicity $d_M$ of
  each   $M$-level. Note that the energies are rescaled in each
  sector   of total spin $S$.}
\label{F-H}
\end{figure}

Nevertheless, it turns out that the special frustration of the
icosahedron leads to different (quasi) degeneracies in several
Hilbert subspaces with total spin S. This can be seen in
\figref{F-H} where the low-lying levels of the icosahedron with
$s=1/2$ are displayed. Note that for better visibility the
energies are rescaled in each sector of total spin $S$. The
numbers display the multiplicities $d_M$ of the
$M$-levels. These are the relevant degeneracies in an applied
field. The total degeneragy at $B=0$ is $d=d_M\cdot
(2S+1)$. Thus, we find also for the icosahedron, that at certain
magnetic field values, not necessarily at the saturation field,
highly degenerate levels will cross and give rise to notable
$(T=0)$--entropy. This can for instance be observed for the
isentrope with $S=2.01 k_B$, which is displayed by the
dashed-dotted line in \figref{F-E}. It approaches the field axis
very closely at about $0.9 \Bsat$.

The unusually large degeneracy in many sectors of total spin is
also the reason for the interesting property that the isentropes
remain at rather small temperatures for decreasing magnetic
fields. This is in strong contrast to spin rings and reflects
the fact that populating the degenerate levels produces
sufficient entropy without increasing the temperature.  The
upturn at small fields close to $B=0$ is mainly due to the
non-degenerate ground state in the sector with $S=0$ which is
separated by large gaps from excited states.

\begin{figure}[ht!]
\centering
\includegraphics[clip,width=85mm]{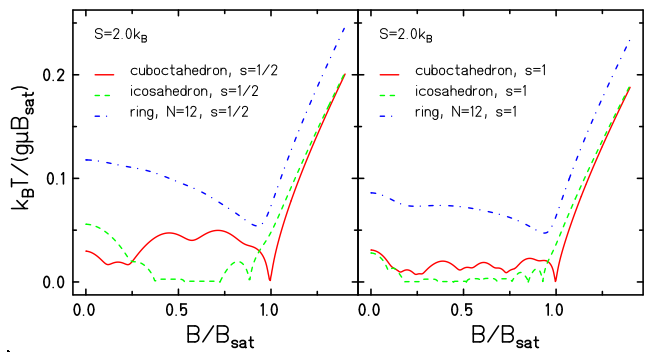}
\caption{(Color online) Isentropes with $S=2 k_B$ of the cuboctahedron, the
  icosahedron, and the ring with $N=12$ with spins $s=1/2$
  (l.h.s.) and $s=1$ (r.h.s.).}
\label{F-F}
\end{figure}

The two panels of \figref{F-F} summarize the above discussion
for the three systems with $N=12$ by comparing the isentropes
with $S=2 k_B$. One sees that at low temperatures close to the
saturation field the cuboctahedron indeed achieves the largest
cooling rate. In the extreme quantum case, i.e. for $s=1/2$ it
also outperforms the two other systems when looking at the
achievable temperatures for $B\rightarrow 0$.  Nevertheless, for
the icosahedron similarly large cooling rates can be achieved
due to a different kind of geometric frustration. For larger
spin quantum numbers, i.e. for becoming more classical, the
differences between the cuboctahedron and the icosahedron tend
to disappear. The unfrustrated ring systems always shows poorer
cooling.

\begin{figure}[ht!]
\centering
\includegraphics[clip,width=85mm]{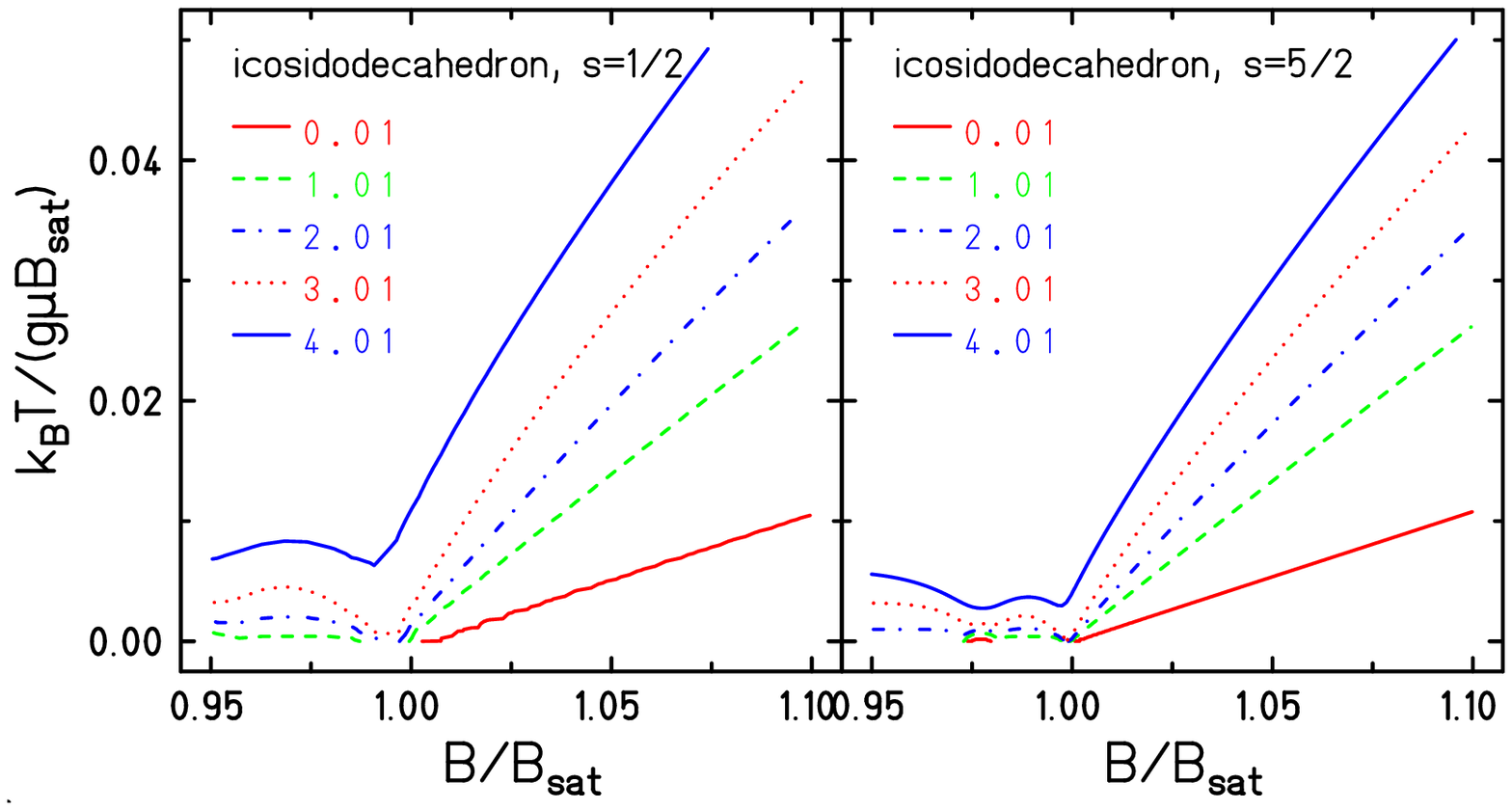}
\caption{(Color online) Isentropes of the icosidodecahedron with
  $s=1/2$ and $s=5/2$. $\Bsat=(12 s |J|)/(g\mu_B)$.} 
\label{F-G}
\end{figure}

Finally we like to discuss the behavior of the antiferromagnetic
icosidodecahedron which also possess icosahedral symmetry. This
Archimedian solid is closely related to the cuboctahedron. It
also consists of corner-sharing triangles, and each spin has
four nearest neighbors.\cite{SSR:EPJB01,mathworld} But compared
to the cuboctahedron it has a much bigger ground-state
degeneracy of 38 at the saturation field, again independent of
spin quantum number.  Figure~\xref{F-G} shows some isentropes of
the icosidodecahedron for the two experimentally relevant cases
of $s=1/2$ and $s=5/2$. Due to the much larger Hilbert space
these isentropes can be evaluated exactly only close to the
saturation field since there only some small subspaces
contribute. The magnetothermal behavior is very similar to the
cuboctahedron with the noticeable difference that now isentropes
with entropies up to $S_0=S(T=0,B=\Bsat)=k_B \ln(38)\approx 3.63
k_B$ head towards $T=0$ at the saturation field.

\section{Summary}
\label{sec-4}

In summary we can say that the investigated frustrated
antiferromagnetic bodies show an enhanced cooling rate in
comparison to non-frustrated (bipartite) spin rings. This rate
is especially large for those systems that show a large
degeneracy of levels, either at the saturation field
(cuboctahedron) or elsewhere (icosahedron).

A few words seem to be in order regarding the question how
realistic the outlined scenario is. In realistic systems the
perfect degeneracy of levels at the saturation field will
certainly be lifted, thus keeping the entropy $S(T=0,B=\Bsat)$
at a small value. Nevertheless, the low-energy density of states
will remain large in the vicinity of the saturation field, since
the originally degenerate levels move not too far, thus the
magnetothermal properties will be left qualitatively unchanged,
see also the studies in Refs.~\onlinecite{DeR:PRB04}.

Finally we like to mention that similar effects can be observed
in interacting electron systems described by the Hubbard
model. This is related to the appearance of flat bands in these
systems.\cite{Tas:PRL92,Mie:JPA92B,HoR:CMP05,DHR:07}

\section*{Acknowledgments}

J.R. and R.S. greatly acknowledge the use of
\verb§spinpack§, a software provided by J\"org
Schulenburg.


\end{document}